\documentclass[iop]{emulateapj}
\def\kms {{\rm km s$^{-1}$}}
\def\msun {\rm {$M_{\odot}$}}
\def\myr {\rm {M$_{\odot} ~yr^{-1}$}}
\def\mdot {$\dot{\rm M}_{\rm acc}$}

\def\ecs {\rm {erg~cm$^{-2}$~s$^{-1}$}}

\def\water {{\rm H$_2$O}}
\def\oh  {\rm OH}
\def\h   {\rm H}
\def\hh   {\rm H$_2$}
\def\o   {\rm O}

\def\co  {\rm CO} 

\shorttitle{Water depletion in Herbig AeBe disks}
\shortauthors{Fedele et al.}
\begin{document}
\title{WATER DEPLETION IN THE DISK ATMOSPHERE OF HERBIG AeBe STARS$^{\star}$}

\author{
D. Fedele\altaffilmark{1}, 
I. Pascucci\altaffilmark{1,2},
S. Brittain\altaffilmark{3},
I. Kamp\altaffilmark{4},
P. Woitke\altaffilmark{5,6,7},
J. P. Williams\altaffilmark{8},
W. R. F. Dent\altaffilmark{9},
W.-F. Thi\altaffilmark{10}}

\altaffiltext{1}{Department of Physics and Astronomy, Johns Hopkins University, 3400 North Charles Street, Baltimore, MD 21218, USA}
  \email{dfedele@pha.jhu.edu}
\altaffiltext{2}{Space Telescope Science Institute, 3700 San Martin Drive,
  Baltimore, MD 21218, USA}
\altaffiltext{3}{Department of Physics and Astronomy, Clemson University, Clemson, SC 29634, USA}
\altaffiltext{4}{Kapteyn Astronomical Institute, Postbus 800, 9700 AV Groningen, The Netherlands}
\altaffiltext{5}{University of Vienna, Dept. of Astronomy,  T{\"u}rkenschanzstr.~17, A-1180 Vienna, Austria}
\altaffiltext{6}{SUPA, School of Physics \& Astronomy, University of St.~Andrews, North Haugh, St.~Andrews KY16 9SS, UK}
\altaffiltext{7}{UK Astronomy Technology Centre, Royal Observatory, Edinburgh, Blackford Hill, Edinburgh EH9 3HJ, UK}
\altaffiltext{8}{Institute for Astronomy, University of Hawaii, 2680 Woodlawn Drive, Honolulu, HI 96822, USA}
\altaffiltext{9}{ALMA JAO, Santiago, Chile}
\altaffiltext{10}{UJF-Grenoble 1 / CNRS-INSU, Institut de Planétologie et d’Astrophysique de Grenoble (IPAG) UMR 5274, Grenoble, F-38041, France}

\begin{abstract}
We present high resolution (R$\sim$100,000) L-band spectroscopy of 11 Herbig
AeBe stars with circumstellar disks. The observations were obtained
with the VLT/CRIRES to detect hot water and hydroxyl radical emission lines
previously detected in disks around T Tauri stars.  \oh\, emission
lines are detected towards 4 disks. The \oh\, $^2\Pi_{3/2}$ P4.5
(1+,1-) doublet is spectrally resolved as well as the velocity profile
of each component of the doublet. Its characteristic double-peak
profile demonstrates that the gas is in Keplerian rotation and points
to an emitting region extending out to $\sim$ 15--30\,AU. 
The \oh\, emission correlates with disk geometry as it is mostly
detected towards flaring disks. None of the Herbig stars analyzed
here show evidence of hot water vapor at a sensitivity similar to that
of the \oh\, lines. The non-detection of hot water vapor emission
indicates that the atmosphere of disks around Herbig AeBe stars are
depleted of water molecules. Assuming LTE and optically thin emission
we derive a lower limit to the \oh/\water\, column density ratio $> 1
- 25$ in contrast to T Tauri disks for which the column density ratio is 0.3
-- 0.4.
\end{abstract}

\keywords{astrochemistry -- molecular processes -- protoplanetary
  disks -- stars: pre-main sequence, Herbig AeBe}

  \section{Introduction}
  Young pre-main-sequence stars are surrounded by gas-rich dust disks
  that are the left over from  the collapse of the molecular cloud core. 
  Sub-micron sized dust grains grow to larger sizes with time
  \citep[e.g.][]{bouwman01, rodmann06}. This leads to the formation of
  planetesimals and eventually planets
  \citep[e.g.][]{weidenschilling93, henning06, blum08}. The evolution of the
  infrared excess in pre-main-sequence stars (PMSs) suggests
  that most of the small dust grains in the disk disappear within 3--5
  Myr from
 the collapse of the molecular cloud and very little
  amount of dust is found beyond 10 Myr
  \citep[e.g.][]{haisch01,bouwman06,hernandez07,roccatagliata09,pascucci10}. A
  similar timescale is found for the evolution of mass accretion rates
  in disks \citep[e.g.][]{mohanty05, jay06, sicilia06, ingleby09, fedele10}.
  The energetic radiation field of a PMS star impinging onto the disk surface
  regulates the disk temperature, can ionize the gas, breaks molecular bonds and
  might lead to evaporation of the outer layers of the disk. An important role
  for the formation of a planetary system as well as for the origin of
  life on Earth is played by water. Compared to other volatiles
  (e.g. {\rm NH$_3$, CO$_2$, CH$_4$}) water has an higher condensation
  temperature. Thus, within a protoplanetary disk, water is the first
  volatile to condense as
 temperature decreases as a function of the
  radial distance from the
 star and vertical depth. In view of the
  cosmic abundance of hydrogen
 and oxygen, water is the most
  abundant ice. This sets a boundary
 ({\it snow line}) beyond which
  most of the molecular
 gas condenses onto ice. For a review on the
  snow line in the solar
 nebula and in protoplanetary disks see,
  e.g., \citet[][]{podolak10}. Beyond the snow
 line the solid
  surface density, and perhaps even the total surface density of the
  disk, increases \citep{kennedy08}. This, in turn, speeds up the
  formation of gas giant
 planets allowing them to form before the
  gas in the disk is
 dispersed. 

 \begin{deluxetable*}{lllllll}
\tabletypesize{\textsize}
\tablewidth{0pt}
\tablecaption{CRIRES observation log \label{tab:log}}
\tablehead{
\colhead{Star} & 
\colhead{RA (J2000)} & 
\colhead{DEC (J2000)} &
\colhead{Date observation} & 
\colhead{Exposure (s)} &
\colhead{Airmass} & 
\colhead{Calibrator} 
}
\startdata
UX Ori    & 05:04:29.9 & -03:47:11.1 & 2008-12-06; 02:50:50  & 1440  & 1.2 & HD 36512 \\
HD 34282  & 05:16:00.5 & -09:48:31.2 & 2008-12-06; 01:47:50  & 1440  & 1.1 & HD 40494 \\
CO Ori    & 05:27:37.8 & +11:25:33.3 & 2008-12-07; 04:32:54  & 720   & 1.2 & HD 64503 \\
V380 Ori  & 05:36:25.0 & -06:43:00.3 & 2008 12 05; 02:36:08  & 720   & 1.4 & HD 23466 \\
BF Ori    & 05:37:12.9 & -06:35:07.5 & 2008-12-07; 05:15:18  & 2400  & 1.1 & HD 64503 \\
HD 250550 & 06:01:59.5 & +16:30:50.7 & 2008-12-06; 05:34:49  & 1080  & 1.3 & HD 41753 \\
HD 45677  & 06:28:17.3 & -13:03:18.2 & 2008-12-06; 07:31:57  & 1080  & 1.1 & HD 74195 \\
HD 259431 & 06:33:04.6 & +10:19:16.6 & 2008-12-06; 06:22:57  & 1080  & 1.2 & HD 74280 \\
HD 76534  & 08:55:08.6 & -43:28:01.1 & 2008-12-05; 07:04:53  & 2400  & 1.1 & HD 28873 \\
HD 85567  & 09:50:28.2 & -60:58:02.9 & 2008-12-06; 08:22:59  & 480   & 1.3 & HD 98718 \\
HD 98922  & 11:22:31.0 & -53:22:07.9 & 2008-12-05; 08:19:43  & 720   & 1.3 & HD 39764 \\	
\enddata
\end{deluxetable*}

  \smallskip
  \noindent
  Water and hydroxyl (\oh) emission have been detected in a number of
  protoplanetary disks around young sun-like stars in the
  near-infrared \citep{carr04, thi05, salyk08, mandell08} as well at
  mid-infrared wavelengths (10 -- 40\,\micron) \citep{carr08, salyk08,
    najita10, pontoppidan10} and in the far-infrared with the {\it
    Herschel Space Observatory} \citep{sturm10, vankempen10}. Lines at
  different wavelengths trace different temperatures and hence
  different regions of the disk. While the near- to mid-infrared and
  high excitation far-infrared lines probe
 the molecular gas from
  the warm surface layers of the disk, the low
 excitation
  far-infrared lines are sensitive to colder gas (a few
 100\,K)
  located closer to the disk midplane \citep[e.g.][]{woitke09}.

  \smallskip
  \noindent
  \citet{mandell08} performed high resolution ($R\sim$27,000)  L-band
  spectroscopy towards two Herbig AeBe stars (AB Aur and MWC
  758). They did not find evidence of hot water vapor emission in
  these two disks. \citet{pontoppidan10} searched for colder molecular
  emission in the mid-infrared the {\it Spitzer Space Telescope}. In
  contrast to T Tauri stars, the mid-infrared spectrum of Herbig AeBe
  stars is poor in molecular emission and they find only tentative
  \water~and \oh~emission lines towards some Herbig AeBe stars.

  \smallskip
  \noindent
  The different molecular emission between T Tauri and Herbig AeBe
  stars might be due to photochemistry which controls the excitation
  and dissociation of molecules. Aiming to test this
  hypothesis, we performed a deep search for hot water and hydroxyl
  radical vapor emission in Herbig AeBe stars with circumstellar
  disks. We performed ultra high resolution ($R \sim$100,000,
  $\Delta$v $\sim$ 3\,\kms) L-band spectroscopy of 11 Herbig AeBe
  stars with CRIRES at the VLT. This spectral resolution allows us to
  resolve the velocity profile of the molecular emission lines and
  thus determine the radii over which the emission arises.

  \noindent
  Observations and data reduction procedures are presented in
  Sec.~\ref{sec:observations}. In Sec.~\ref{sec:results} we describe the
  immediate results of the survey: the detection of the \oh\,$^2\Pi_{3/2}$
  P4.5 (1+,1-) rovibrational line and the non-detection of water rovibrational
  lines. In Sec.~\ref{sec:analysis} we present the analysis of the \oh\,
  lines and discuss the implications of our findings in
  Sec. \ref{sec:discussion}. Finally, we summarize our results in
  Sec. \ref{sec:conclusions}.

  \section{Observations and data reduction}\label{sec:observations}
  High resolution L-band spectroscopy was obtained on the nights
  of 4, 5 and 6 December 2008 with the ESO's VLT cryogenic
  high-resolution infrared echelle spectrograph
  \citep[CRIRES,][]{kaufl04} at the Paranal observatory in Chile. A
  slit width of 0\farcs2 was adopted. In order to center the target
  inside the narrow slit we used the adaptive optics system and the
  targets themselves as reference stars. This procedure reduces slit
  losses resulting in a higher signal-to-noise ratio and higher
  spectral resolution. Unresolved sky emission lines can be used to
  determine the actual spectral resolution. The
  full-width-half-maximum (FWHM) of the \oh~sky lines measured in the
  raw frames is $\sim 3$\,\kms (or $R \sim$100,000) as expected from the
  nominal resolution. The spectra were recorded by nodding the
  telescope along the direction perpendicular to the slit (oriented
  along the parallactic angle) with a nodding throw of 10\arcsec. The
  spectra cover the wavelength range between 2861 $-$ 2936\,nm (order
  19, $\lambda_{ref}$ = 2909.6\,nm) thus covering several
  ro-vibrational lines of \water\, detected in comets
  \citep[e.g.][]{dellorusso04, dellorusso05} and in protoplanetary
  disks \citep{salyk08} as well as ro-vibrational lines of \oh\, detected
  in disks by \citet{mandell08} and \citet{salyk08}.

  \noindent
  The data were reduced using the ESO CRIRES pipeline
  v.11.0\footnote{http://www.eso.org/sci/data-processing/software/pipelines/index.html}
  following a standard approach: first the CRIRES frames are corrected for
  flat field, dark and bad pixels; second the frames at a given nodding
  position are combined together and the two master frames (one for
  each nodding position) are then combined together after correcting for
  the spatial offsets (due to the nodding procedure); finally, the spectrum is
  extracted using a rectangular mask. The sky emission lines are used to
  determine the wavelength dispersion solution. To properly correct for
  atmospheric telluric absorptions the spectra were bracketed with
  observations of standard stars of early spectral type
  (see Table~\ref{tab:log}). These spectra are divided by the stellar
  atmosphere models of \citet{kurucz79} to determine the instrument
  response function and atmospheric transmission curve. 
  The optical depth of the telluric lines is scaled to the depth of the science
  target. Finally, each science spectrum is divided by the response
  function. The spectral regions heavily absorbed (atmospheric
  transmission $<$ 0.5) are not used in the rest of the analysis. 
  For the flux calibration we scaled the CRIRES spectra to the observed L-band
  magnitude measured by \citet[][V380 Ori, HD 76534, UX Ori, HD 250550,
    HD 259431, HD 45677, BF Ori]{dewinter01}, \citet[][HD 98922, HD
    34282, HD 85567]{malfait98} and \citet[][CO Ori]{morel78}. 
  
  \smallskip
  \noindent
  Given the high temperatures of the stellar photosphere of Herbig AeBe
  stars, the L--band spectrum is relatively free of stellar photospheric absorption
  lines. The spectra of all the stars observed with CRIRES are continuum
  dominated and have high signal-to-noise ratio (S/N $>$ 100). Such an
  high ratio allows to detect weak emission lines down to a few percent
  of the continuum flux level. The exposure times vary from object to
  object and range from $\sim$ 10\,min (for targets of L $\sim 4 -
  6$\,mag) to 40\,min (L $\sim 7$\,mag). The log of the observations is
  given in Table \ref{tab:log}.

\begin{deluxetable*}{lllllllll}
\tabletypesize{\footnotesize}
\tablewidth{0pt}
\tablecaption{Properties of the program stars \label{tab:stars}}
\tablehead{
\colhead{Star}    & 
\colhead{Sp.Type} & 
\colhead{Age}     & 
\colhead{A$_{\rm V}$}   &
\colhead{D}       &
\colhead{L$_{\rm UV}$\tablenotemark{*}} & 
\colhead{\mdot} &
\colhead{Group} \\
 & & [Myr] & [mag] & [pc] & [L$_{\odot}$] & [10$^{-7}$\,\myr] &  }
\startdata
UX Ori  & A3 & 4.5\tablenotemark{a}      & 0.8\tablenotemark{c} & 460\tablenotemark{g} & 2.4  & 0.66\tablenotemark{i}&II\\
HD 34282& A0 & 7\tablenotemark{b}        & 0.6\tablenotemark{c} & 160\tablenotemark{e} & --   & $<$0.2\tablenotemark{j}& I \\
CO Ori  & F7 & $<0.12$\tablenotemark{c}  & 2.2\tablenotemark{c} & 460\tablenotemark{g} & 2.9\tablenotemark{**} &0.9\tablenotemark{k} &II\\
V380 Ori  & A1 & $<$0.01\tablenotemark{c}&1.43\tablenotemark{e} & 430\tablenotemark{g} & 18.8 & 25\tablenotemark{i}  & I \\  
BF Ori    & A5 &3.2\tablenotemark{a}     & 0.37\tablenotemark{f}-0.7\tablenotemark{c} & 460\tablenotemark{g} & 0.6-1.3  & 0.87\tablenotemark{i}&II\\
HD 250550 & B7 & 0.25\tablenotemark{c}   & 1.17\tablenotemark{c}& 700\tablenotemark{g} & 160  & 0.16\tablenotemark{i}&I \\
HD 45677  & B2 &--                       & &$>$500\tablenotemark{h}                    & $>$65 & &II\\
HD 259431 & B5 & $<$0.01\tablenotemark{c}&0.63\tablenotemark{f}-1.62\tablenotemark{e} & 800\tablenotemark{g} & 225-2200  &  7.8\tablenotemark{i}&I \\
HD 76534  & B2 & $\gtrsim$\,0.5\tablenotemark{d}& 0.80\tablenotemark{f}&870\tablenotemark{g}   & 1020\tablenotemark{***}  & &I \\
HD 85567  & B5 & $<$0.01\tablenotemark{c}&2.23\tablenotemark{c} & 1500\tablenotemark{c}& --    & &II\\
HD 98922  & B9 & $<$0.01\tablenotemark{c}&0.54\tablenotemark{c} & 1000\tablenotemark{c}& 545   & 17.4\tablenotemark{j}& II\\      
\enddata
\tablenotetext{*}{Integrated between 1100 - 2430\,\AA\,, ({\it IUE}
  spectrum, \citealt{valenti03}) and multiplied by a factor 10$^{0.4 \cdot
  A_{1770}}$. Where A$_{1770}$ is the extinction at $\lambda$
= 1770\,\AA~ measured from A$_{\rm V}$ using the R = 3.1 extinction
relation of \citet{cardelli89}.}
\tablenotetext{**}{Missing short wavelength {\it IUE} spectrum,
  integrated between 1850 - 2430\,\AA.}
\tablenotetext{***}{Missing long wavelength {\it IUE} spectrum,
  integrated between 1150 - 2000\,\AA.}
\tablenotetext{a}{\citet{montesinos09};}
\tablenotetext{b}{\citet{merin04};}
\tablenotetext{c}{\citet{manoj06};}
\tablenotetext{d}{\citet{martinzaidi08};}
\tablenotetext{e}{\citet{vandenancker98};}
\tablenotetext{f}{\citet{valenti03};}
\tablenotetext{g}{\citet{hillenbrand92};}
\tablenotetext{h}{\citet{dewinter97};}
\tablenotetext{i}{Donehew et al. submitted;}
\tablenotetext{j}{\citet{garcialopez06};}
\tablenotetext{k}{\citet{calvet04};}
\end{deluxetable*}

\begin{figure*}
\centering
\includegraphics[width=10cm, angle=90]{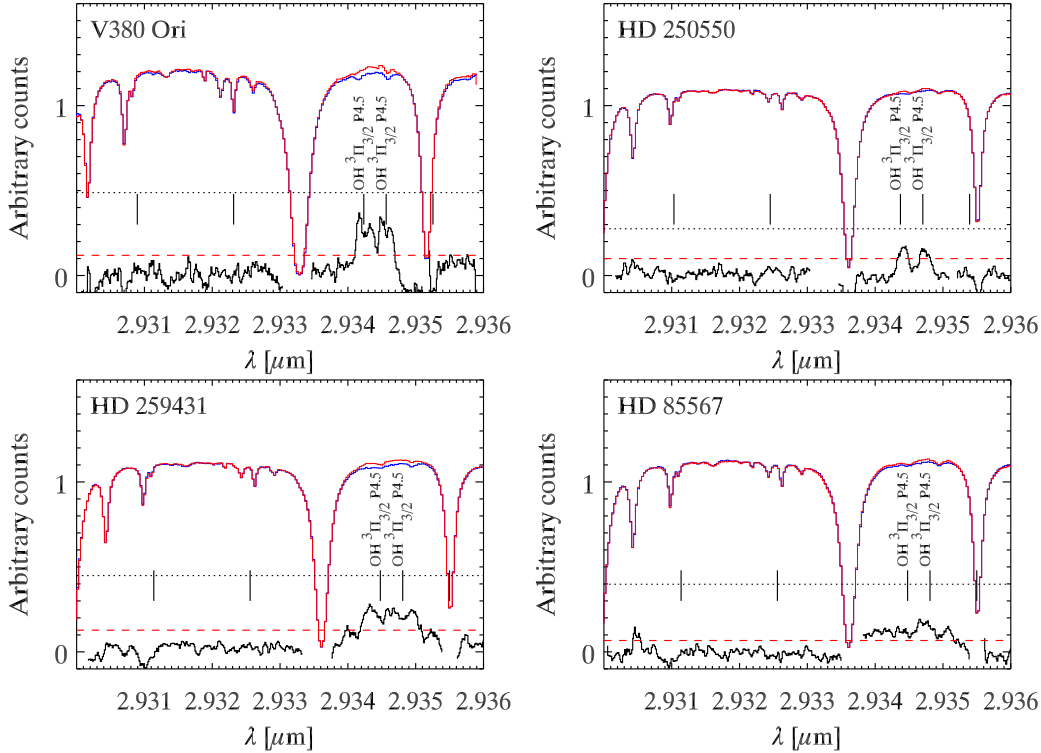}
\caption{Portion of the CRIRES spectra for V380 Ori, HD 250550, HD
  259431 and HD 85567. In each panel the upper line (red) is the
  extracted target spectrum along with the atmospheric transmission
  curve (blue line). The lower plot of each frame is the spectrum
  cleaned by the telluric absorption lines multiplied by a factor of
  10. The dashed line shows the $3\sigma$ detection level. For each
  star the \oh\, P4.5 (1+,1-) doublet is detected and spectrally
  resolved. The vertical lines show the positions of some of the
  strongest \water\, rovibrational lines. The dotted line indicates
  the expected line height of the \water\, (001--000) [11$_{66}$ --
    12$_{67}$] transition in the case of \oh/\water\, column density
  ratio equal to unity. \label{fig:oh}}
\end{figure*}

\begin{figure}
\centering
\includegraphics[width=8cm]{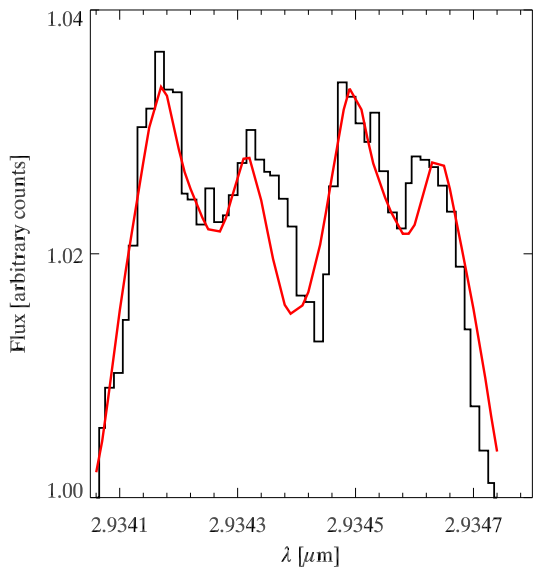}
\caption{\oh\, P4.5 doublet towards V380 Ori. The red line is the best
fit model (Sec.~\ref{sec:analysis_ohline}, Table~\ref{tab:model})
assuming Keplerian rotation. The asymmetry of the line profile
indicates a deviation from the Keplerian motion of the gas
(see Appendix). \label{fig:oh1}}
\end{figure}

\begin{figure}
\centering
\includegraphics[width=8cm]{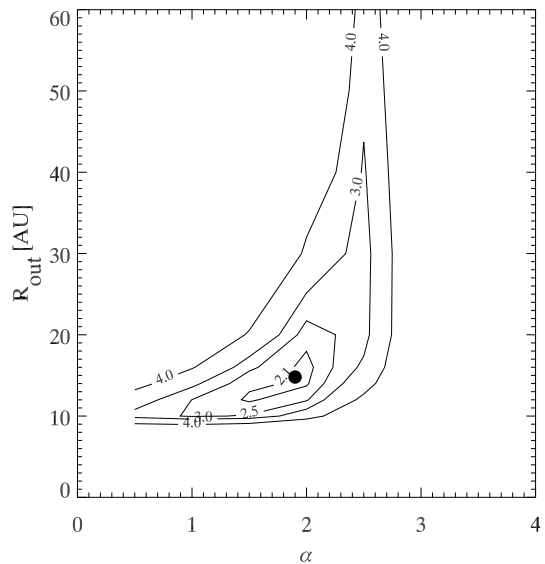}
\caption{$\tilde{\chi}^2$ contours in the $\alpha - {\rm R_o}$
  parameter space for V380 Ori assuming i = 30$^{\circ}$ and R$_{\rm
    i} = 2\,$AU. The best fit value is shown as a black
  dot. \label{fig:chi2_1}}
\end{figure}

\begin{figure}
\centering
\includegraphics[width=8cm]{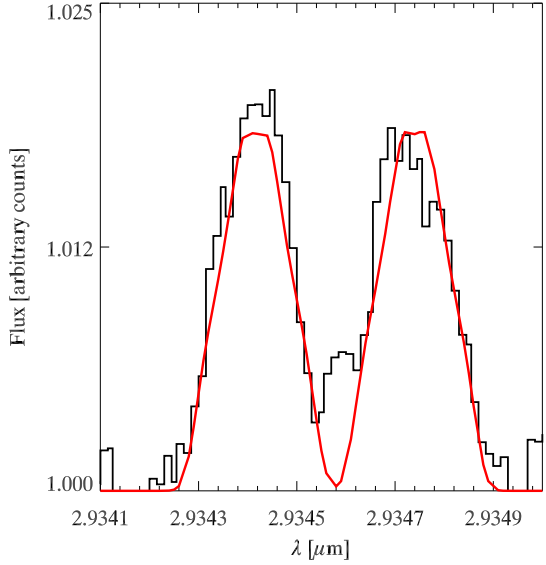}
\caption{Same as Figure~\ref{fig:oh1} for HD 250550 \label{fig:oh2}}
\end{figure}

\begin{figure}
\centering
\includegraphics[width=8cm]{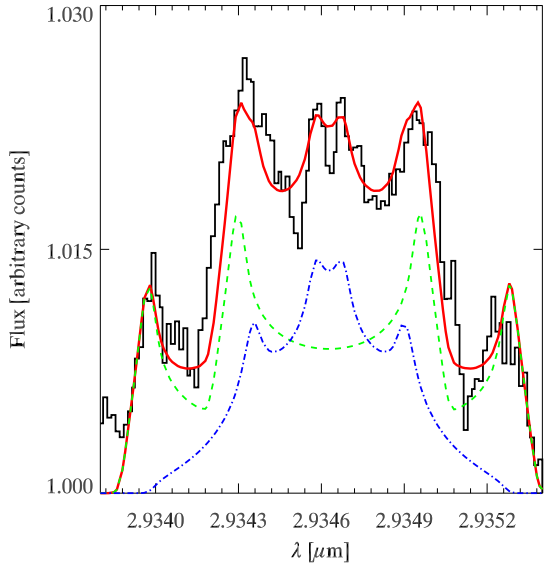}
\caption{Same as Fig.~\ref{fig:oh1} for HD 259431. The dashed and dot-dashed
  lines are the the optically thick and disk's atmosphere components
  respectively. 
 \label{fig:oh3}}
\end{figure}

\begin{figure}
\centering
\includegraphics[width=8cm]{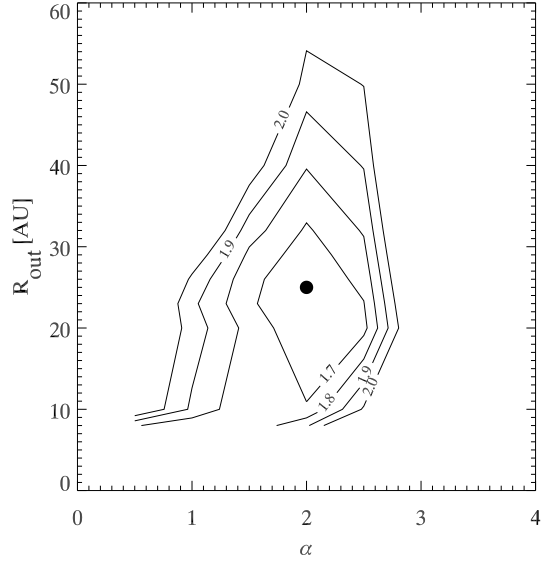}
\caption{$\tilde{\chi}^2$ contours in the $\alpha_2 - {\rm R_{2,o}}$
  parameter space for HD 259431 assuming i = 50$^{\circ}$,
  $\alpha_{\rm 1} = 0$, R$_{\rm 1,i} = 0.9\,$AU and R$_{\rm 1,o}$ =
  R$_{\rm 2,i} = 1.4\,$AU. The best fit value is shown as a black dot.
 \label{fig:chi2_2}}
\end{figure}

\begin{figure}
\centering
\includegraphics[width=8cm]{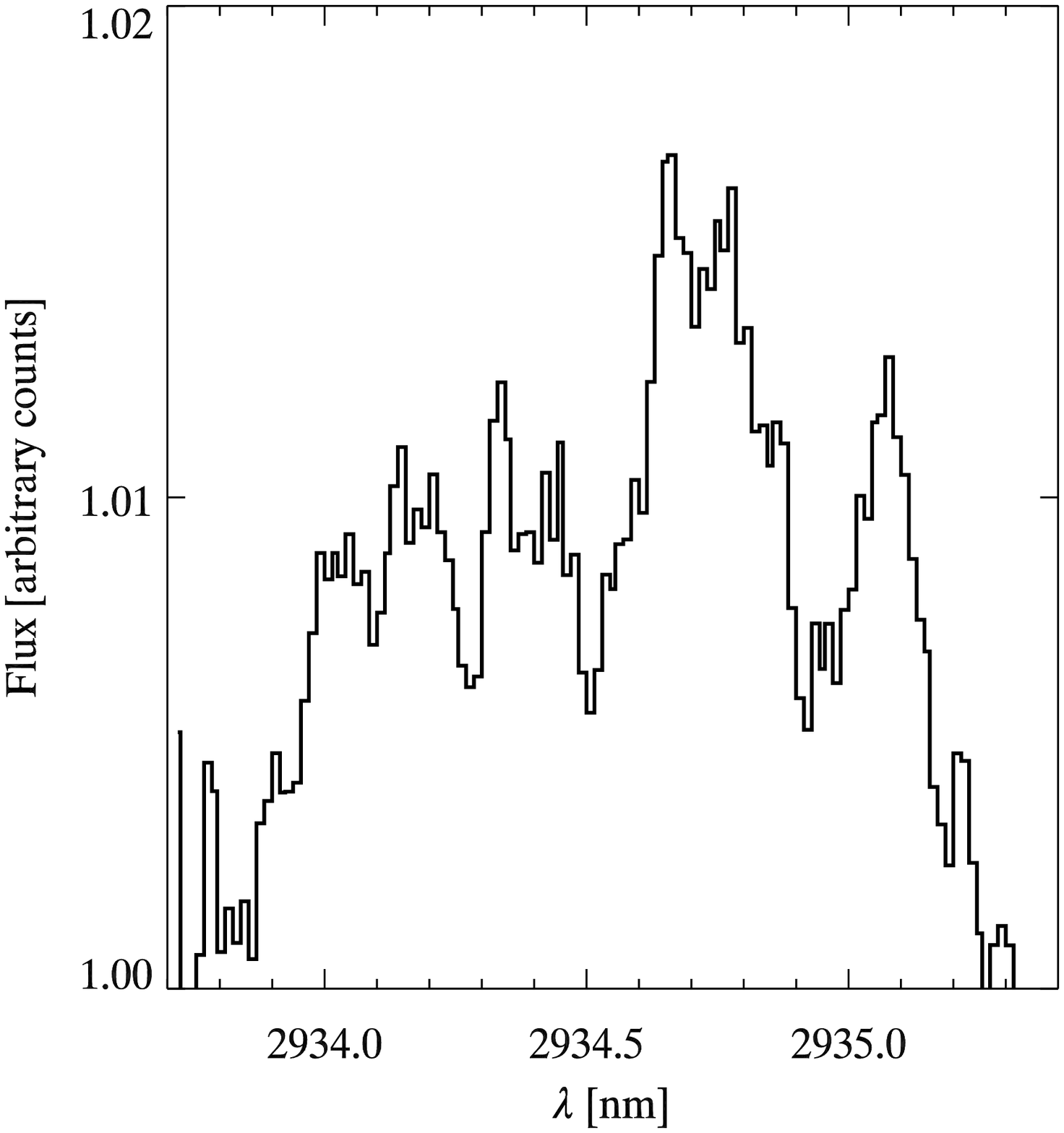}
\caption{\oh\, P4.5 doublet towards HD 85567. 
 \label{fig:oh4}}
\end{figure}

 \section{Immediate results}\label{sec:results}
  We have searched for \water~and \oh~rovibrational lines previously
  detected in protoplanetary disks around sun-like stars, comets, as
  well as transitions that are not yet detected but reported in recent
  synthetic molecular line lists. In the following section we summarize
  the main findings for \water~and \oh. 
  
  \subsection{Detection of \oh\, lines}
  The rovibrational \oh\, $^2\Pi_{3/2}$ P4.5 (1+,1-) at
  2.9345\micron\, transition is detected
 towards 4 disks (V380 Ori,
  HD 250550, HD 259431 and HD 85567,
 Fig.~\ref{fig:oh}). The high
  spectral resolution of CRIRES allows us
 to resolve not only the
  doublet but also the velocity profile of
 each transition. Modeling
  of the line profiles and estimates of the
 radii traced by this
  emission are presented in
 Sec. \ref{sec:analysis}.
  
  \noindent
  The profile of the $^2\Pi_{3/2}$ P4.5 (1+,1-) line varies from object
  to object (Figs.~\ref{fig:oh1}-\ref{fig:oh3}). 
  In the case of V380 Ori each line shows a characteristic double-peak
  profile as one would expect if the gas is in Keplerian rotation. 
  In the case of HD 250550 the lines are narrower 
  reflecting the lower inclination of the disk
  (Table\,\ref{tab:model}). Finally, the \oh\, line in HD 259431 and HD
  85567 is very broad and the two transitions of the doublet are
  blended even at the high resolution of CRIRES. In the case of HD
  259431, multiple peaks are visible hinting to multiple velocity
  components. Two peaks are detected (above $3\sigma$) at the sides of
  the line at 2.9340\,\micron\, and 2.9352\,\micron\, respectively. The
  position of these peaks is symmetric to the center of the emission
  and correspond to a velocity of $\pm$~47\,\kms. The \oh\,
  emission is very broad (FWHM = 90\,\kms) and the sides of the
  emission might be contaminated by the strong telluric absorptions at
  2.9335\,\micron\, and 2.9355. We used different telluric standard
  stars (taken at different airmass) to check the reliability of the
  line profile. In all cases we found the two high velocity wings
  above the $3\sigma$ level. Further confirmation of the presence of
  these wings comes from {\it a posteriori} analysis of the line
  profile (Sec.~\ref{sec:analysis_ohline}): if the \oh\, emission is
  made of a single velocity component, the central part of the
  emission would be stronger than the sides due to the blending of the
  line (see e.g. dotted-dashed line in Fig.~\ref{fig:oh3}). The
  analysis of the line profile confirms the presence of multiple
  velocity components and the reliability of the high velocity
  peaks. At higher velocity, where the atmospheric transmission
  function drops below 70$\%$, the emission is heavily corrupted by
  telluric absoprtion hence we excluded these regions from the
  analysis of the line. 

  \noindent
  Within the spectral range covered by our observations there
  are other two transitions of the \oh~$^2\Pi_{3/2}$ P branch (J = 2.5,
  3,5) as well as two transitions of the \oh~$^2\Pi_{1/2}$ P branch (J =
  2.5, 3.5). Unfortunately all four transitions fall either in highly
  saturated telluric absorption regions or inside the inter-chips gaps
  hence we cannot compute stringent line flux upper limits.

  \subsubsection{\oh\, column density}\label{sec:results_oh}
  The energy level of the ro-vibrational transitions of \oh~might be
  populated either thermally or radiatively (fluorescence). In the
  first case the collisions with atoms and molecules are the dominant
  excitation mechanism and the emission is only dependent on the
  temperature and density of the gas as we will show below (LTE case). 
  In the case of fluorescent emission the ro-vibrational levels are
  populated by absorption of near-infrared photons (in the electronic
  ground state) or absorption of UV photons which excite the
  electronic states with following cascade in the electronic ground
  state. UV fluorescent pumping in protoplanetary disks has been
  investigated for the first time by \citet{brittain03, brittain07} to
  explain the \co~emission from the disk around the HAeBe stars HD
  141569 A and AB Aur. \citet{mandell08} implemented UV and IR
  fluorescent pumping for the \oh~ro-vibrational lines detected in the
  disk around AB Aur and MWC 758. They find that a fluorescent model
  can explain the observed \oh~line intensities as well as \oh~thermal
  emission. The difference between thermal and fluorescent emission is
  in the inferred \oh~column densities, which are higher in the LTE
  case. Since our observations do not allow us to discriminate between
  LTE and fluorescent emission, we assume in the following analysis
  that the \oh~gas is in LTE, the emission is optically thin, and
  compute the total number of \oh\, molecules from the measured line
  flux. For the gas temperature we take a value of 700\,K as estimated
  by \citet{mandell08} for the Herbig AeBe stars AB Aur and MWC
  758. With these assumptions the \oh\, line intensity is
  \citep{herzberg50}

  \begin{equation}\label{eq:oh_density_1}
    {\rm I}(\oh) = N_n(\oh) h {\tilde \nu} c A_{nm}
  \end{equation}

  \noindent
  with {\rm I(OH) the \oh\, line luminosity (= 4 {\rm $\pi$ d$^2$ F$_{\oh}$) with
    d being the distance to the star and F$_{\oh}$ the measured \oh\,
    flux), {\rm N$_{\rm n}$} the number of molecules in the upper state {\rm n},
    {\rm h} Planck constant, {\rm c} the light velocity, {\rm {$\tilde
        \nu$}} the wavenumber of the transition ({\rm cm$^{-1}$}), and {\rm
      A$_{nm}$} the  Einstein coefficient of the $^2\Pi_{3/2}$ P4.5
    transition. In LTE the quantity {\rm N$_{\rm n}$} relates to the total number
    of \oh\, molecules ({\rm N}) as 

  \begin{equation}\label{eq:oh_density_2}   
    \frac{{\rm N_n(OH)}}{{\rm N(OH)}} = \frac{{\rm g_J~exp(-E_n/k T)}}{{\rm Q(T)}}
  \end{equation}

  \noindent
  where {\rm g$_{\rm J}$ (\,=\,2J+1)} is the degeneracy of the state, {\rm E$_{\rm n}$} the
  energy of the upper state, {\rm k} the Boltzmann constant and {\rm Q(T)} the
  partition sum. Substituting {\rm N$_{\rm n}$} from Eq.~\ref{eq:oh_density_2}
  in Eq.~\ref{eq:oh_density_1} we have

  \begin{equation}\label{eq:ohdensity}
   {\rm N(OH)} = \frac{{\rm I(OH) Q(T) ~exp(E_n/k T)}}{{\rm hc {\tilde \nu} A_{nm}(2J + 1)}} 
  \end{equation} 

  The transition parameters and the partition sum are taken from
  HITRAN \citep{rothman09}. The total number of molecules can be
  converted into a vertical column density ({\rm N$_{{\rm col}}$}) by dividing
  it for the area of the \oh\, emitting region, $\pi$(R${\rm _o^2}$ -
  R${\rm _i^2}$), with R$_{\rm i}$ and R$_{\rm o}$ being inner and
  outer radius of the \oh\, emitting region (see
  Sec.~\ref{sec:analysis_ohline}). When the \oh\, P4.5 doublet is not detected
  and the emitting region is not known we assume R${\rm _i}$ = 0.4\,AU and
  R${\rm _o}$ = 10\,AU and a line width of 30~\kms\, (based on our results
  of the \oh\, line detections). The values of N and N$_{{\rm col}}$
  are listed in Table~\ref{tab:result}. We note that increasing the
  temperature to T(OH) = 1000\,K decreases N (hence N$_{{\rm
      col}}$) almost by an order of magnitude.

  \subsection{Non-detection of \water~lines}
  In order to search for faint water vapor emission we first created a
  line list using the water line list catalog of \citet{barber06}. The
  latter is the most complete and accurate water line list in the
  literature. We used their {\it BT2} code to produce the synthetic
  spectrum and adopted the same temperature as that for the \oh\, gas
  (700\,K). Some of the strongest water lines detected by
  \citet{salyk08} in the T Tauri stars AS 205 A and DR Tau lie around
  2930\,nm close to the \oh\, $^2\Pi_{3/2}$ P4.5 doublet we have detected
  in 4 disks and hence in a clean part of the spectrum. For comparison
  Fig. \ref{fig:oh} shows the positions of the strongest water
  emission lines from \citet{salyk08}. None of the 11 Herbig AeBe
  stars observed with CRIRES show evidence of hot water vapor emission
  in the L-band spectra.
  
  \smallskip
  \noindent
  We compute the 3-$\sigma$ upper limit of the line flux for the line
  \water~(001--000) [11$_{66}$ -- 12$_{67}$] at 2.931~\micron\,
  detected in T Tauri stars. The upper limit is measured as the
  product $3 \times \sigma
 \times \Delta\lambda$, where $\sigma$ is
  the standard deviation of the
 spectrum and $\Delta\lambda$ is a
  charateristic line width of 30\,\kms or
  2.9~$\times$~10$^{-4}$\micron~ (of the same order of the \oh\, line
  width). The line flux upper limits
 for the program stars are given
  in Table \ref{tab:result}. Typical
 upper limits are of the order
  of $10^{-16} - 10^{-15}$\,\ecs. The water lines detected by
  \citet{salyk08}\footnote{Estimated from the
  Fig. 2 of
    \citet{salyk08}}
 have fluxes of the order of $10^{-15} -
  10^{-14}\,$\ecs. 

  \subsubsection{\water~column density}\label{sec:h2o_density}
  With the same assumptions of Sec.~\ref{sec:results_oh} (optically
  thin emission, LTE, {\rm T} = 700\,K) we can estimate the upper limit to
  the total number of molecules and column density of \water. In the
  case of water, Eq.~\ref{eq:ohdensity} must be modified to take into
  account an additional degeneracy associated with the nuclear spin
  ({\rm g$_{\rm I}$} = 3 for {\rm J} odd and {\rm g$_{\rm I}$} = 1 for
  {\rm J} even). We compute the upper limits of {\rm N}(\water) and
  {\rm N$_{{\rm col}}$}(\water) using the line flux upper limits of
  the (001--000) [11$_{66}$ -- 12$_{67}$] transition. The results are
  listed in Table~\ref{tab:result}. We note that also in this case the
  number of molecules decreases by almost an order of magnitude by
  increasing the temperature to T(H${\rm _2}$O) = 1000\,K.

  \subsection{\oh/\water\, line ratio}
  From the comparison of Fig.~\ref{fig:oh} with Fig.~2 of \citet{salyk08}
  it is evident that the \oh/\water\, line ratio is higher in Herbig
  AeBe disks than in T Tauri disks. For the Herbig stars with detected
  \oh\, emission lines, we have that the \oh/\water\, line flux
  ratio is $\ga 1.5 - 30$\footnote{We refer here to the total flux of
    the \oh\, P4.5 doublet, i.e. twice the value listed in the fourth
    column of Table~\ref{tab:result}}. This translates into a lower
  limit to the \oh/\water\, column density ratio of 1 -- 25
  (Table~\ref{tab:result}). For comparison we show in
  Fig.~\ref{fig:oh} the expected line height of the \water\,
  (001--000) [11$_{66}$ -- 12$_{67}$] transition in the case of
  \oh/\water\, column density ratio equal to unity (dotted line). The
  intensity of the \oh\, P4.5 line for the four Herbig AeBe stars
  (Table~\ref{tab:result}) is of the order of $10^{-15} -
  10^{-14}$\,erg~cm$^{-2}$~s$^{-1}$, similar to the \oh\, and \water\,
  line
 flux measured in T Tauri disks \citep{salyk08}. However, in
  the case of T Tauri disks the \oh/\water\, line flux ratio is $\sim$
  1 and the column density ratio is 0.3 -- 0.4, that is, the
  \oh/\water\, column density ratio is $\sim 3 - 25$ times larger in
  Herbig AeBe disks than in T Tauri disks. This suggest that, in
  contrast to T Tauri disks, water vapor is less abundant than \oh\,
  in the disk atmosphere of Herbig AeBe stars.

  \section{Analysis}\label{sec:analysis}
  In order to understand the pattern of detections and non-detections
  we investigate whether there is any correlation between stellar and
  disk properties and the surveyed emission lines
  (Sec.~\ref{sec:analysis_geometry}). We also model the profiles of
  the resolved \oh~lines and provide  estimates for
  the radial distances traced by the L-band \oh\, emission in disks
  around Herbig AeBe stars (Sec.~\ref{sec:analysis_ohline}).

 \begin{deluxetable}{lllll}
\tablewidth{0pt}
\tablecaption{Stellar and disk parameters \label{tab:model}.}
\tablehead{& & V380 Ori & HD 250550 & HD 259431\tablenotemark{*}}
\startdata
M$_*$     &  [\msun]     & 2.8\tablenotemark{a} & 3.6\tablenotemark{b} & 6.6\tablenotemark{c}\\
i         &  [$^{\circ}$] & 30 -- 35  & 8 -- 15 & 50\tablenotemark{c}\\
$\alpha$&              & 2.0 & 2.5 & 0 \\
${\rm R_i}$  &  [AU]       & 2.0 -- 2.1 & 0.4 -- 0.7 & 0.8 -- 1.0  \\
${\rm R_o}$  &   [AU]      & 12 -- 16  & 100\tablenotemark{d} & 1.3 -- 1.5  \\
$\alpha_2$&          & & & 2.0                  \\
${\rm R_{2,i}}$  &  [AU]   & &  & 1.3 -- 1.5\tablenotemark{e}  \\
${\rm R_{2,o}}$  &  [AU]    & &  & 23 -- 30     \\
$\tilde{\chi}^2$        & & 1.9    & 1.2  & 1.4 \\
\enddata
\tablenotetext{*}{In the case of HD 259431 the model is the sum of two
  velocity components. The subscript ``2'' refers to the second velocity component}
\tablenotetext{a}{\citet{hubrig09};}
\tablenotetext{b}{\citet{hernandez04};}
\tablenotetext{c}{\citet{kraus08};}
\tablenotetext{d}{Fixed;}
\tablenotetext{e}{Constrained to be $\ge {\rm R_i,o}$}
\end{deluxetable}

 \subsection{\oh\, lines and disk properties}\label{sec:analysis_geometry}
  A noticeable distinction appears when detections are compared with the disk
  geometry: the majority of the sources with detection of \oh~emission
  appears to have a flared disk geometry. \citet{meeus01} have shown that the
  ratio of the far-infrared to the near-infrared flux is sensitive to the
  geometry of the disk. Far-infrared bright stars belonging to the group I of
  \citet{meeus01} are thought to have a flared disk geometry, that is the disk
  scale height ({\rm H}) and opening angle ({\rm H/R}) of the disk
  increases with the distance from the star ({\rm R}). Group II
  sources are thought to have a flat, self-shadowed geometry, due to
  the disk inner rim casting a shadow at larger disk radii
  \citep{dullemond04}. \citet{boekel03} found that the two groups
  occupy two different regions in the IRAS $m_{{\rm 12}} - m_{{\rm
      60}}$ color versus $L_{{\rm NIR}}/L_{{\rm IR}}$ diagram, where
  $L_{{\rm NIR}}$ is the integrated luminosity as measured by the J,
  H, K, L and M photometry and L$_{{\rm IR}}$ is the integrated
  luminosity measured by the IRAS 12, 25 and 60\,\micron
  ~fluxes. Following \citet{boekel03} we provide in
  Table~\ref{tab:stars} the group of the program stars. Interestingly,
  three out of four stars for which \oh~emission is detected (V380
  Ori, HD 250550, HD 259431) belong to group I while only one source
  showing \oh~emission (HD 85567) belongs to group II. We note that
  the other two HAeBe stars with detected \oh~emission in the
  literature \citep[AU Aur and MWC 758][]{mandell08} also belong to
  group I. We discuss the origin of such a correlation in
  Sec.~\ref{sec:discussion_geometry}.

  \subsection{\oh\, line profiles}\label{sec:analysis_ohline}
  In this section we present an analysis of the \oh\, P4.5 (1+,1-) line
  profiles for V380 Ori, HD 250550, HD 259431 and HD 85567. 
  Figs. \ref{fig:oh1}-\ref{fig:oh3} show the velocity profile of the
  doublet for the 4 stars from which we can estimate the radial
  distribution of the molecular gas. We assume that the line intensity
  follows a power-law distribution as a function of the radial
  distance from the star \citep[e.g.][]{smak81,carmona07,vanderplas09}
  of the form
  
  \begin{equation}\label{eq:intensity}
    {\rm I_{OH}(R)} = {\rm I_{OH}(R_0) \cdot (R/R_0)^{-\alpha}}
  \end{equation}
  
  \noindent
  where {\rm R} is the distance from the star and {\rm I$_{{\rm OH}}$(R$_0$)}
  is the intensity at the inner radius. We also assume that the \oh~line is
  optically thin. If the line is
  optically thick we should include in the analysis the continuum and line
  optical depths \citep[e.g.][]{horne86b}. Here, we are interested in
  determining the inner and outer radius of the \oh~emitting region and a
  complete analysis of the line emission properties is beyond the scope of
  this paper. The radial profile of the \oh~line is converted into a velocity
  profile assuming that the gas is in Keplerian rotation. In addition the
  model line is convolved with a velocity width {\rm v }= $\sqrt{{\rm
      v_{in}}^2 + {\rm v_{th}}^2}$ where {\rm v$_{\rm in}$} is the
  instrumental broadening ($\sim$~3\,\kms) and {\rm v}$_{\rm th}$~(=
  $\sqrt{{\rm 2kT/m_{\oh}}}$) is the thermal broadening of the line
  (where {\rm k} is the Boltzmann constant, {\rm T} is the gas
  temperature and {\rm m}$_{\oh}$ is the mass of \oh). We assume a
  temperature of 700\,K as found by \citet{mandell08} which corresponds
  to {\rm v}$_{th} \sim$ 0.8\,\kms.  For the Keplerian rotation and
  line convolution we used the IDL  codes ``keprot''  and
  ``convolve''\footnote{http:$\/\/$www.ster.kuleuven.ac.be$\/\sim$bram$\/$dark$\_$theme$\/$work.html}
  described in \citet{acke05}.

  \smallskip
  \noindent
  With the assumption of Keplerian rotation and of the radial profile 
  in Eq.~\ref{eq:intensity}, the velocity profile of the
  emission line is a function of 4 parameters: the power law intensity
  exponent $(\alpha)$, the disk inclination (${\rm i}$), the inner
  and outer radii of the \oh\, emitting region (${\rm R_i, R_o}$). The
  Keplerian velocity is ${\rm v_{kep} = \sqrt{GM_*/R}}$, hence the inner
  radius determines the maximum velocity in the line profile.  Because
  the disk inclination is not known for our sources (apart from HD
  259431) we fitted the \oh~P4.5 doublet by minimizing the reduced
  $\chi^2$ ($\tilde{\chi}^2$) between the
 observed and model velocity profile

  \begin{equation}\label{eq:chisquared}
    \tilde{\chi}^2 = \frac{1}{\rm d}\sum \Big ( \frac{\rm model - observed}{\sigma} \Big )^2
      \end{equation}

  \noindent
  with d degree of freedom. In our fitting procedure we
 leave free
  only three parameters: ${\rm i}$, ${\rm R_i}$ and ${\rm R_o}$ and
  repeat the model fitting for five different values of $\alpha$ (0,
  1.0, 1.5, 2, 2.5). The best fit parameters are listed in
  Table~\ref{tab:model}. 
  
  \subsubsection{Determination of the outer disk (R$_{\rm o}$)}
  The measure of R$_{\rm o}$ requires further explanation. The
  contribution from the outer disk to the line intensity is reduced
  because of the power-law distribution (Eq.\ref{eq:intensity}).
  Moreover, the fit of the line profile might be degenerate if the
  parameters are not independent. As we saw above, ${\rm
    R_i}$ is uniquely determined by the maximum velocity of the
  line. In a Keplerian motion, the disk inclination and outer radius
  are responsible of the characteristic double-peak profile: for a
  given ring of gas at a distance ${\rm R}$ from the central star the
  two peaks of the line are separated by the quantity ${\rm v_{kep}
    \cdot sin(i) \propto R^{-1/2} \cdot sin(i)}$. If the disk
  inclination is known, it is possible to unambiguously determine
  ${\rm R_o}$. The power-law index ($\alpha$) controls how steeply the
  intensity intensity decreases with radius (hence velocity) and
  affects only slightly the peak-to-peak separation as shown in Fig.~1
  of \citet{smak81}. To check the robustness of the fit we first
  determine the best fit parameters, then we fix the disk inclination
  and inner radius and compute the $\tilde{\chi}^2$ surface on the
  $\alpha - {\rm R_o}$ parameter space. 

  \smallskip
  Notes on individual targets:
  
  \smallskip
  \noindent
  {\it V380 Ori.}
  The \oh\, emission extends from $\sim$ 2\,AU to $\sim$ 15
  from the central star (Fig.~\ref{fig:oh1}) and we estimate a disk
  inclination of $\sim$ 30$^{\circ}$. Fig.~\ref{fig:chi2_1} shows
  the $\tilde{\chi}^2$ of the fit in the $\alpha - {\rm R_o}$ plane. The
  single peaked $\tilde{\chi}^2$ distribution shows that the determination of
  ${\rm R_o}$ is robust. The \oh~P4.5 line profile is
  asymmetric with the blue peak stronger than the red peak. The model in
  Fig. \ref{fig:oh1} has been manually modified to account for the
  excess flux in the blue-shifted peak by multiplying the latter by a
  factor of 1.2. The origin of the asymmetric line profile is
  discussed in the Appendix.

  \noindent
  {\it HD 250550.}
  In this case the line is top-flat and the two peaks are not
  visible. This is likely due to the low inclination of the disk in
  the plane of the sky (disk almost face-on). In this situation it is not
  possible to determine unambiguously ${\rm R_o}$ so we fix it to a
  value of 100\,AU. The best fit is found for an inner radius of
  $\sim$ 0.4\,AU and an inclination of 10$^{\circ}$.

  \noindent
  {\it HD 259431.}
  The \oh\, line profile in this case is more complicated and appears to have
  multiple components. We note that we were not able to fit the \oh\,
  line profile with a model made of a single velocity component as in
  the previous cases. Multi-wavelength interferometric observations reveal
  the presence of optically thick gas within the dust sublimation radius
  \citep{kraus08}. In particular, the near-infrared continuum is dominated by
  optically thick gas that is accreting onto the star, while the mid-infrared
  continuum arises from the passively irradiated disk atmosphere at larger
  radii. Similarly, the analysis of the \hh\, FUV lines \citep{bouret03}
  suggests that the molecular hydrogen spectrum (seen in absorption) has
  multi-temperature components. In particular they find that an hot component
  (T $\sim$ 1300\,K) comes from optically thick gas in the vicinity ($>$ 0.5
  AU) of the star. This is in good agreement with the detection of the high
  velocity wings in the \oh\, line profile (Sec.~\ref{sec:results}).   
  To account for the presence of optically thick gas inside
  the dust truncation radius we modify the standard model including a second
  component (see Fig.~\ref{fig:oh3}). Thus the free parameters of the fit are
  now 5: disk inclination, inner and outer radius of the optically
  thick component (R$_{\rm 1,i}$, R$_{\rm 1,o}$), inner and outer
  radius of the passively irradiated disk component (R$_{\rm 2,i}$,
  R$_{\rm 2,o}$). We adopt a disk inclination of 50$^{\circ}$ based on
  the results of \citet{kraus08}. For the thermal broadening of the line
  we assume T$_1$=1300\,K (= T(H$_{\rm 2}$)) and {\rm T$_2 = 700$\,K}
  \citep[from][]{mandell08} for the high and low velocity component
  respectively. Finally, we assume a constant radial dependence of the
  line intensity from the radius ($\alpha = 0$ in
  Eq. \ref{eq:intensity}) for the high velocity component. Our best
  model parameters are in good agreements with the results of the
  interferometric observations: the first component extends from
  $\sim$ 0.8\,AU to $\sim$ 1.4\,AU and the second components extends
  from $\sim$ 1.4\,AU to $\sim$ 25\,AU from the star. We note that our
  fitting routine tends to produce a low value of ${\rm R_{2,i}}$
  (lower than ${\rm R_{1,o}}$), thus we constrain ${\rm R_{2,i}}$ to
  be $\ga {\rm R_{1,o}}$. Fig.~\ref{fig:chi2_2} shows the $\tilde{\chi}^2$ of
  the fit in the $\alpha - {\rm R_{2,o}}$ plane (after fixing all the
  other parameters). The value of ${\rm R_{2,o}}$ is not degenerate with
  $\alpha$. 

  \noindent
  Recently \citet{bagnoli10} found a similar result for the
  [\ion{O}{1}] line at 6300\,\AA. High spectral resolution observations
  revealed the presence of an high velocity component between $\sim$ 1
  -- 2\,AU and a low velocity component peaking around
  20\,AU. Compared to the [\ion{O}{1}] line profile, the \oh~high
  velocity component is clearly double-peaked and allows us to
  constrain better the outer radius of the high velocity
  component (R$_{\rm 1,o}$). Compared to the [\ion{O}{1}] 6300\,\AA ~line,
  the \oh\, emission detected here likely arises from deeper layers
  (higher A$_{\rm V}$) in the disk. 

   \noindent
  {\it HD 85567.}
  The \oh\, emission is broad (of the order of $\sim$ 100\,\kms) and
  suggests the presence of high velocity molecular gas. As in the case
  of HD 259431, the high velocity \oh~emission might originate in an
  optically thick gas inside the dust sublimation radius. There are no
  estimates of the disk inclination in the literature and given the
  low signal-to-noise of the spectrum it is hard to estimate the
  radial extent of the \oh~emitting region. Further observations are needed
  to measure the properties of the \oh~emission from this disk.

  \section{Discussion}\label{sec:discussion}
  The main findings of Sec.~\ref{sec:results} \& Sec.~\ref{sec:analysis} are:
  i) the non-detection of water vapor emission lines, ii) the
  detection of \oh~emission lines from the disk atmosphere of Herbig
  AeBe stars and iii) the correlation between OH emission and disk
  flaring. In contrast, observations of T Tauri stars have shown that
  there exist a hot (500--1000\,K) water layer in their disk
  atmosphere with densities similar to the OH layer we detect in
  Herbig AeBe stars \citep[e.g][]{carr08,salyk08}.
  In this section we discuss the processes affecting the formation and
  destruction of water molecules in disks. The discussion is
  structured as follow: first we explain how water molecules form in
  the disk atmosphere (Sec.~\ref{sec:discussion_formation}); then we
  address the water (and \oh) destruction processes and outline the
  main differences between T Tauri and Herbig AeBe stars
  (Sec.~\ref{sec:discussion_destruction}). Finally we examine how
  non-stationary processes might affect the water content in different
  regions of the disk (Sec.~\ref{sec:discussion_transport} and
  \ref{sec:discussion_decoupling}) with a final discussion on the
  origin of the \oh\, disk geometry correlation
  (Sec.~\ref{sec:discussion_geometry}).

  \subsection{Formation of water in disks}\label{sec:discussion_formation}
  The formation of water molecules in the warm disk atmosphere is a
  three steps process

  \begin{eqnarray}
    \h + \h + dust & \rightarrow & {\rm H_2} + {\rm dust} \label{eq:water_I_a}\\
    \o + {\rm H_2} & \rightarrow & \oh + \h \label{eq:water_I_b} \\ 
    \oh + {\rm H_2} &  \rightarrow & {\rm H_2O} + \h \label{eq:water_I_c}
  \end{eqnarray}

  \noindent
  The formation of \hh~on warm dust grains in the disk atmosphere
  (Eq.~\ref{eq:water_I_a}) is justified by the findings of
  \citet{cazaux02} who measure a moderate \hh\, formation rate on warm
  (up to 900\,K) dust grains. Further \hh\, molecules might form
  through gas phase reactions \citep[e.g.][]{glassgold09} 
  
  \begin{equation}\label{eq:h2}
    {\rm H^- + H} \rightarrow {\rm H_2} + {\rm e}
  \end{equation}
  
  \noindent
 The neutral-neutral reactions
  (Eqs.~\ref{eq:water_I_b} \& \ref{eq:water_I_c}) require high
  temperature to occur (T $> 300$\,K). If \hh\, is absent and hydrogen is
  mainly atomic, water may form through radiative association reactions
  (e.g. Kamp et al. submitted)
  such as 
  
  \begin{eqnarray}
    \h + \o  & \rightarrow & {\rm OH} + h\nu        \label{eq:water_I_d}\\
    \oh + \h & \rightarrow & {\rm H_2O} + h\nu \label{eq:water_I_e}
  \end{eqnarray}
  
  \noindent
  In both cases (neutral-neutral or radiative association reactions)
  the gas phase formation of water in the disk atmosphere strongly
  depends on the gas density. 
  \smallskip
  \noindent

  Recently a number of stationary disk chemical models have shown that water
  can be efficiently formed in the warm disk surface
  layer. \citet{glassgold09} (hereafter G09) investigated {\it in situ}
  water formation with a model based on X-ray (only) heating and
  ionization of the disk atmosphere which is most appropriate for T
  Tauri stars. \citet{woitke09} (hereafter
  ProDiMo model) compute the thermo-chemical structure of Herbig AeBe
  disks which generally have L${\rm _X}/$L$_{\rm FUV} << 1$ \citep{kamp08} using
  only UV/optical irradiation. According to their model there exists a
  hot water layer at distances between 1 -- 30\,AU and relative height
  z/r $\la 0.1-0.3$ where water molecules are thermally decoupled from
  the dust (T(H${\rm _2}$O) $>$ T(dust)). In this model
  \oh~(and \co) are abundant within 30\,AU from the star and above an
  A$_{\rm V}$ of a few where water is not efficiently formed. Water is only
  found in deeper layers where it is shielded from photodissociation
  and densities/temperatures are high enough to form it through
  neutral-neutral gas-phase chemistry. 
   
  \subsection{Destruction of water in disks}\label{sec:discussion_destruction}
  Pre-main-sequence stars emit strong ultraviolet radiation. For
  classical T Tauri stars UV photons are produced by three main
  components: the stellar photosphere, the enhanced chromospheric
  activity, and the magnetospheric accretion. In an X-ray irradiated T
  Tauri disk (as the one investigated by G09) water molecules are
  dissociated in the disk atmosphere through charge transfer reaction
  with {\rm  H$^+$}. In the case of early spectral type stars (B and
  A) as the ones studied here, the stellar photosphere is the major
  source of FUV radiation and overwhelms any contribution from
  mass accretion or stellar activity. The UV radiation field impinging
  onto the surface of the circumstellar disk affects the physical
  properties (such as the gas and dust temperature) of the disk as
  well as its chemistry. In the case of
 the \oh/\water~chemistry,
  the strength of the UV field regulates the
 formation/destruction
  rate. Due to the strong soft UV radiation water can be easily
  photodissociated \citep[e.g.][]{woitke09}

  \begin{eqnarray}%\label{eq:water}
    {\rm H_2O} + h\nu & \rightarrow & \oh + \h \label{eq:water_II_a} \\
    & \rightarrow & \o + 2\h \label{eq:water_II_b} \\
    & \rightarrow & \o + {\rm H_2} \label{eq:water_II_c} 
  \end{eqnarray}

  Water can also be destroyed by charge transfer
  reactions as in the case of T Tauri stars (e.g. G09). However, given
  the intense ultraviolet radiation, UV photodissociation is likely to
  be the major dissociation mechanism in the atmosphere of an Herbig
  Ae/Be star as we will show next.
 
  \smallskip
  \noindent
  The minimum energy required to dissociate water molecules is 5.1 eV
  \citep[e.g.][]{harich00} which corresponds to a radiation of  $\lambda$ =
  2430\,\AA. The photodissociation cross section ($k_{pd}$) of water
  varies with the color impinging radiation field. As an example, the
  cross section for an incoming radiation of a 10,000\,K blackbody is
  nearly 4 times larger that for a 4,000\,K blackbody
  \citep{vandishoeck08}. This is due to the stronger soft UV radiation
  (relevant for the photodissociation of water molecules) of Herbig
  AeBe stars compared to T Tauri stars. Column 4 of
  Table~\ref{tab:stars} lists the UV luminosity of the program stars
  integrated between 1100\,\AA ~and 2430\,\AA. The integrated
  luminosity is multiplied by a factor of $10^{0.4 \cdot A_{1770}}$ to
  correct for interstellar extinction. The quantity $A_{1770}$ is the
  extinction at $\lambda = 1770\,$\AA ~and it is measured from A${\rm _V}$
  using the R = 3.1 extinction relation of \citet{cardelli89}. The
  UV luminosity of the program stars ranges between $\sim 1 -
  10^3$\,L$_{\odot}$. In the case of a classical T Tauri star, the UV
  luminosity ranges between $\sim 0.01 - 1$\,L$_{\odot}$. In particular,
  for the two stars in \citet{salyk08} (AS 205A, DR Tau) L$_{UV}$ is
  $\sim 0.1 - 1$\,L$_{\odot}$. Thus, the UV radiation emitted by the
  Herbig AeBe stars in this sample is up to 4 orders of magnitude
  larger than that in the T Tauri stars studied by \citet{salyk08}.
  Photodissociation is a plausible mechanim to explain the
  lack of hot water vapor lines in Herbig AeBe disks. 
  
  \noindent
  A way to further test this scenario would be to detect \oh\,
  ro-vibrational lines in the mid-infrared at high J rotational
  levels. In fact \oh~molecules formed by the photodissociation of
  water (eq.~\ref{eq:water_II_a}) are vibrationally and rotationally
  excited \citep[e.g.][]{dutuit85,vanharrevelt00,harich00}. As a
  consequence, the high rotational levels of \oh~are easily populated
  \citep[e.g.][]{carrington64, harich00, bonev06} and
  mid-infrared spectra could be able to detect the high J-value
  transitions such as found in the outflow of HH 211 \citep{tappe09}
  and in TW Hya \citep{najita10}. A few high J \oh~lines have been
  detected in the mid-infrared towards Herbig AeBe stars with disks
  (Sturm, private communication). We note however that, depending on
  the density of the environment, the \oh\, molecules might be
  thermalized very fast. This might prevent us from detecting any
  trace of water photodissociation. In this regard, the rotational
  diagram of L-band \oh\, lines in the two stars studied by
  \citet{mandell08} is characterized by a single rotational
  temperature.  
   
  \noindent
  In addition to photodissociation there are effects that could prevent us
  from detecting water vapor emission from the atmosphere of Herbig
  AeBe disks. The first effect to consider is the temperature
  difference between \oh\, and \water\, lines detected here and in
  \citet{salyk08}. The lower energy state of the \oh\, P4.5 doublet is
  $\sim$ 500\,K while the energy state of \water\, transitions covered
  by our observations are $\ga$ 1000\,K. Thus the \oh\,emission
  potentially traces colder gas than the \water\,emission. One might
  speculate whether the \water\, non-detections are due to the
  different temperature layers probed
 in the disk. We find this
  possibility unlikely given that the \oh\,
 transitions detected by
  \citet{mandell08} in AB Aur and MWC 758 have
 energies up to $\sim$
  2360\,K (\oh\, P9.5), similar to the \water\, detected by
  \citet{salyk08}. Thus the high \oh/\water\,line flux ratio
 in
  Herbig AeBe stars is a signature of water depletion in the disk
  atmosphere rather than of transitions tracing different
  temperatures. 
    
 \smallskip
  \noindent
  Another argument that is often used to explain the paucity of molecular
  emission lines in Herbig AeBe disks is that the strong infrared excess
  might be able to veil the faint emission of molecular lines in the infrared
  \citep[e.g.][]{pontoppidan10}. In this case the molecular emission lines are
  veiled by the dust continuum. In this regard we note that the we detect
  \oh~emission at similar wavelengths and similar sensitivity of the
  \water~ro-vibrational transitions detected in T Tauri disks. Thus, if water
  vapor is present, it must be located deeper in the disk (at higher {\rm
    A$_V$}) where is colder than the \oh~detected here and/or thermally
  coupled with the dust (i.e. T(H${\rm _2}$O) = T(dust)).

  \subsubsection{\oh~photodissociation}
  The most probable photodissociation channel of water is the one that
  produces an \oh~ molecule \citep[$\sim$ 80\% -  90\%,
    e.g.][]{crovisier89,combi04,harich00} that is followed by 

  \begin{equation}\label{eq:oh}
    \oh + h\nu  \rightarrow  O + H 
  \end{equation}

  \noindent
  The photodissociation cross section of \water\, is only twice as
  large as that of the \oh\, for an incoming blackbody radiation of
  10,000\,K which is representative for an Herbig Ae star
  \citep{vandishoeck08}. In addition, the minimum energy needed to
  break the \oh\, bound is 4.47 eV which corresponds to a radiation of
  2616\,\AA~ \citep{vandishoeck83}, very similar to that of
  water. This suggests that \oh~ molecules should also photodissociate
  as is seen in comets \citep[e.g.][]{combi04}. The dissociation of
  \oh\, by UV photons (Eq.\ref{eq:oh}) is thought to be the main
  reservoir of the excited Oxygen ($^1$D) atoms which produce the
  6300\,\AA ~transition often in solar system comets and in Herbig
  AeBe stars \citep[e.g.][]{vandishoeck84, storzer98,
    morgenthaler01,acke05}\footnote{We note that another source of
    O($^1$D) atoms is the direct photodissociation of water
    (Eq.~\ref{eq:water_II_c})}. The similar radial distributions of
  \oh\, and [\ion{O}{1}] in HD 259431 \citep[Sec.~\ref{sec:analysis}
    and][]{bagnoli10} might be a direct evidence of the formation of
  O($^1$D) atoms via \oh~photodissociation.

  \smallskip
  \noindent
  The previous discussion suggests that the atmosphere of Herbig AeBe
  disks is depleted in water vapor. The detection of \oh\, emission and
  the column density ratio \oh/\water\, $\ga$ 1 suggests that
  \oh\, molecules are produced again {\it in situ}
  (Eq.~\ref{eq:water_I_b}).

  \subsection{Transport of water in disks}\label{sec:discussion_transport}
  Non-stationary processes might affect the content of water in
  disks. Within a protoplanetary disk gas and solid particles can
  migrate either inwards or outwards \citep[for a review
    see][]{ciesla06} or mixed by e.g. turbulent motions. Outwards
  migration was suggested, e.g., by \citet{stevenson88} to explain the
  formation of Jupiter. Inward migration was studied by
  \citet{ciesla06} with the aim at addressing the abundance of hot
  water vapor found in some T Tauri stars. Water can efficiently form
  in the disk interior on the surface of dust grains. If water-rich
  bodies such as icy planetesimals migrate inward and pass the snow
  line, the water ice evaporates from their surface and enrich the
  inner disk of water vapor. For such a process to happen however, the
  migration of planetesimals must be very fast and occur on a
  timescale shorter than the disk lifetime. Testing this hypothesis is
  important to probe the initial conditions of planetary systems. In
  this regards, two parameters are important: 1) the abundance
  of deuterated water and 2) the ortho-para ratio of water
  \citep[e.g.][]{encrenaz08}. Both these parameters are sensitive to
  the temperature formation and hence to the region where water forms
  in the disk (atmosphere or disk interior). We invite the interested
  reader to look into the recent work of \citet{thi10} for a detailed
  theoretical treatment of the formation of deuterated water.   

  \noindent
  The case studied by \citet{ciesla06} is appropriate for T Tauri
  stars and does not include photodissociation. In the case of Herbig
  AeBe stars however, the transport and/or mixing of volatiles might
  induce the depletion of water in the disk interior by dredging up
  molecules to the surface layers where
 molecules are
  photodissociated. Two timescales are important in this
 regard: 1)
  the chemical relaxation timescale ($\tau_{chem}$) which
 regulates
  how fast chemical reactions occur, hence the formation
 of water
  and 2) the mixing timescale. If the vertical mixing is
 faster than
  the chemical relaxation, the disk interior can be
 depleted of
  water. Estimates of $\tau_{chem}$ for an Herbig Ae disk
 computed
  with ProDiMo are $\la 10^2$\,yr. The simulations
 of the transport
  processes in disks by \citet[e.g.][]{ilgner04} suggest that
  vertical mixing is slower and does not affect the global chemical
  evolution of the disk.
  
  \noindent
  In this regard, ProDiMo predicts very different chemical structures
  between T Tauri and Herbig AeBe disks: T Tauri disks are colder and
  $\tau_{chem}$ may be as long as 10$^8$\,yr in the dense mid-plane
  where water is condensed in ice. Disks around Herbig AeBe stars
  instead have hotter and more (chemically) active environment and
  there is only very little amount of icy water in the mid-plane.

  \subsection{Gas-dust decoupling and dust settling}\label{sec:discussion_decoupling}
  An important parameter for the photochemistry of disks is the physical
  decoupling of gas and dust in the disk atmosphere
  \citep[e.g.][]{meijerink09}. Direct observational evidence of gas
  and dust physical decoupling was found for the Herbig Ae stars HD
  101412 \citep{fedele08} and HD 95881 \citep{verhoeff10}. In the case
  of physical decoupling, the disk atmosphere is depleted in large
  dust grains (which settle into the midplane) and water molecules are
  unshielded from the UV radiation. In such a case the molecular
  content in the disk atmosphere might be easily reduced by UV
  photodissociation. The situation changes dramatically if water and
  \oh\, {\it self shielding} is taken into account as in
  \citet{bethell09}. According to their model, in a dust-depleted disk
  atmosphere water and \oh\, molecules become the major source of UV
  opacity and are able to block the photodissociative radiation from
  penetrating further into the disk. As a consequence, the abundance
  of water is enhanced even in the presence of a moderate FUV
  radiation (L$_{\rm FUV} = 10\,$L$_{\rm \sun}$). One problem of the water
  self-shielding model of \citet{bethell09} is that it assumes the
  presence of \hh\, in the dust-depleted disk atmosphere. Without dust
  however, \hh\, form via the ${\rm H^-}$ route which produces an
  overall lower abundance of molecular hydrogen. On the other hands,
  the model of \citet{meijerink09} as well as G09 and ProDiMo do not
  include water self-shielding. This makes difficult to compare the
  observations with predictions from different models at this stage.

  \smallskip
  \noindent
  We do not have direct evidence of physical decoupling in the systems
  studied here but we do have evidence for thermal decoupling. If we
  assume a gas temperature of T $\sim~700\,$K as found by
  \citet[]{mandell08} for AB Aur and MWC 758, the gas is hotter than
  the dust \citep[T(dust) $\la 300\,$K, e.g.][]{kamp04} at a spatial
  scale of $\sim $1 -- 20\,AU, the size of the \oh\, emitting region
  (Sec. \ref{sec:analysis}). Thus, the \oh\, vapor detected here is
  likely thermally decoupled from the dust. 

  \subsection{OH emission and disk geometry}\label{sec:discussion_geometry}
  In Sec.~\ref{sec:analysis_geometry} we found that the presence of
  \oh\, ro-vibrational lines correlates with the geometry of the
  disk: \oh\, emission is mainly detected towards flaring disks (group I). 
  As we saw in Sec.~\ref{sec:results_oh}, the excitation of the \oh\,
  ro-vibrational transitions depends on the temperature and density of
  the gas in the case of thermal emission and on the intensity of the UV
  and near-infrared radiation field in the case of fluorescence. In
  the case of Herbig AeBe stars the disk temperature is regulated by
  the UV output of the star as well as the near-infrared continuum
  emitted by the disk. Thus the intensity of \oh\, ro-vibrational
  lines depends on the UV radiation field impinging into the disk
  surface regardless of the excitation mechanism. In a flaring disk geometry,
  the disk surface area that is illuminated by the central star is
  much larger than in a self-shadowed geometry. In the latter case
  \oh\, might still be excited but the emitting area is small (and the
  emission weak) and confined to the inner part of the disk.  
 
  \smallskip
  \noindent
  A trend is also found between CO fundemantal ro-vibrational emission
  at 4.7\micron~ and disk flaring (van der Plas et al. in
  preparation). CO emission originates from a larger disk area in
  flaring disks compared to flat disks. A larger sample of Herbig AeBe
  disks and more \oh~transitions are necessary to pin down the OH line
  excitation mechanism.
 
  \section{Conclusions}\label{sec:conclusions}
  This paper presents high resolution L-band spectroscopic observations of
  Herbig AeBe stars with disks. Among the 11 stars analyzed here, four
  show hot \oh\, emission which appears to correlate with the disk
  geometry and, in turn, with the UV irradiation impinging on the
  disk. The detection rate of \oh\, emission is 70\% for disks with a
  flared geometry \citep[including the two systems analyzed
    by][]{mandell08} and $\sim$ 13\% (1 out of 6) for self-shadowed
  disks. The \oh\, ro-vibrational lines are spectrally resolved
  allowing us to measure an extension of $\sim$ 10--30\,AU for the
  \oh\, emitting region. In contrast to T Tauri, Herbig AeBe stars do
  not show evidence of hot water vapor in their spectra. The detections of \oh\,
  emission lines and the observed \oh/\water\, column density ratio
  ($\ga$ 1) indicate that the atmosphere of disks around Herbig AeBe
  stars is depleted in water molecules. Given the stronger UV
  radiation field of Herbig AeBe stars compared to T Tauri stars, a
  plausible explanation for the non-detection of water lines is that
  water in the disk atmosphere is dissociated by UV photons.  The
  absence of hot water vapor in the disk atmosphere does not preclude
  the existence of colder water deeper in the disk (at lower
  temperature and higher A$_{\rm V}$). Far-infrared water lines have
  been detected with the {\it Herschel Space Observatory} towards the
  Herbig Ae star HD 100546 \citep{sturm10} (although further
  confirmation is needed). This indicates the presence of warm water
  vapor in the disk.

\bibliographystyle{apj}
\bibliography{ms}

  \begin{appendix}\label{appendix}

    \section{On the asymmetric \oh\, line profile in V380 Ori}
    As we saw in Sec.~\ref{sec:analysis_ohline}, the \oh\, line profile in V380
    Ori is asymmetric, with the blue-shifted component brighter than
    the red-shifted one. A similar asymmetry was found towards other
  protoplanetary disks in the [OI] 6300\,\AA~ line for the Herbig Ae
  star HD 100546 \citep{acke06} and in the \co\, fundamental
  rovibrational lines in EX Lupi \citep{goto11}. Such an
  asymmetry is either due to a deviation from the Keplerian motion of
  the gas or to a non homogeneous distribution of the emitting
  gas. The inner radius of the \oh\, emitting region ($\sim$ 2\,AU) is
  too large  to be coincident with the dust sublimation
  radius. \citet{alecian09} found evidence of a close low-mass star
  companion to V380 Ori. They determine a projected angular separation
  ${\rm a\,sin(i)} < 0.33$\,AU. Regardless of the inclination, the orbit
  of the companion is certainly within the inner rim of the \oh\,
  emitting region. Circumbinary disks have inner gap of the order of
  $\sim$ 2 -- 3 times the binary separation \citep{artymowicz94} which
  brings the truncation radius to $\sim$ 1 -- 2\,AU for V380 Ori
  (depending on the eccentricity, inclination and mass ratio of the
  binary). This is remarkably close to the \oh\, inner radius estimated
  here. The companion might be also the source of the perturbation of
  the gas dynamics seen in the \oh\, line profile. Recent simulation by
  \citet{regaly11} show that in the case of a
  circumbinary disk the velocity distribution of the gas differs from
  the circular Keplerian case. The disk may become eccentric (due to
  tidal interaction with the lower mass companion) and the velocity
  profile of the emerging lines is asymmetric in a fashion similar to
  Fig.~\ref{fig:oh1}. 
  \end{appendix}

\acknowledgments
D. Fedele thanks B. Acke for kindly providing the IDL script
``keprot'' used in the analysis. DF thanks A. Carmona, G, van der
Plas, M. E. van den Ancker, A.M. Mandell, D. Neufeld, C. P. Dullemond
for useful discussion and suggestions. We are grateful to the ESO
staff in Garching and on Paranal for performing the CRIRES
observations in service mode. DF is supported by a NSF Astronomy \&
Astrophysics research grant to IP (ID\#: AST0908479). We thanks the
anonymous referee for the helpful comments.

\facility{VLT, CRIRES}.

  \begin{deluxetable}{lcccccccc}
    \tabletypesize{\scriptsize}
    \tablewidth{\linewidth}
    \tablecaption{\oh\, and \water\, lines parameters\tablenotemark{*} \label{tab:result}}
    \tablehead{
      \colhead{Star}            & 
      \colhead{\oh\, FWHM}         &
      \colhead{\oh\, $\left| {\rm EW} \right|$}        &
      \colhead{L$_{\rm OH}$}    & 
      \colhead{L$_{\rm H_2O}$ @ 2.931\,\micron} & 
      \colhead{log(N(OH))}         &
      \colhead{log(N$_{\rm col}$(OH))}     &
      \colhead{log(N(H$_{\rm 2}$O))}       &
      \colhead{log(N$_{\rm col}$(H$_{\rm 2}$O))}     \\
      & 
              [\kms]                            &
              [10$^{-6}$\micron]                   &
              [10$^{-15}$\,erg~cm$^{-2}$~s$^{-1}$] &
              [10$^{-15}$\,erg~cm$^{-2}$~s$^{-1}$] &
              [molecules]&
              [mol./cm$^{-2}$] &
              [molecules]&
              [mol./cm$^{-2}$]
    }
    \startdata
    UX Ori    &       & & \multicolumn{2}{c}{$<$ 3.0}                      &$<$ 44.3 & $<$ 15.4 & $<$ 44.3 & $<$ 15.5\\  
    HD 34282  &       & & \multicolumn{2}{c}{$<$ 2.0}                      &$<$ 43.2 & $<$ 14.3 & $<$ 43.2 & $<$ 14.4\\  
    CO Ori    &       & & \multicolumn{2}{c}{$<$ 1.0}                      &$<$ 43.8 & $<$ 15.0 & $<$ 43.9 & $<$ 15.0\\   
    V380 Ori  & 27    & 8 & 8.4 ($\pm$ 5.0) & $<$ 4.5                       &    45.0 &     16.1 & $<$ 44.5 & $<$ 15.6\\ 
    BF Ori    &       & & \multicolumn{2}{c}{$<$ 1.0}                      &$<$ 43.8 & $<$ 15.0 & $<$ 43.9 & $<$ 15.0\\
    HD 250550 & 18    & 2.8 & 1.4 ($\pm$ 0.1) & $<$ 2.0                       &    44.6 &     15.9 & $<$ 44.5 & $<$ 15.8\\
    HD 45677  &       & & \multicolumn{2}{c}{$<$ 10}                       &$<$ 43.9 & $<$ 15.0 & $<$ 43.9 & $<$ 15.1\\
    HD 259431 & 45\tablenotemark{**} & 12 & 14.5 ($\pm$ 2)  & $<$ 1.0       &    45.7 &     15.9 & $<$ 44.3 & $<$ 14.5\\
    HD 76534  &     & & \multicolumn{2}{c}{$<$ 1.5}    &$<$ 44.5 & $<$ 15.7 & $<$ 44.6 & $<$ 15.7\\
    HD 85567  & 30\tablenotemark{**} & 8      & 10.0 ($\pm$ 30)  & $<$ 3.0                        &    46.1 &     17.3 & $<$ 45.4 & $<$ 16.5\\
    HD 98922  &     &  & \multicolumn{2}{c}{$<$ 1.2}                      &$<$ 44.6 & $<$ 15.7 & $<$ 44.6 & $<$ 15.8\\
    \enddata
    \tablenotetext{*}{Upper limits are computed as the product $3 \times
      \sigma \times \Delta \lambda$, with $\sigma$ standard deviation
      between 2.930\,\micron\, and 2.936\,\micron, and $\Delta \lambda$
      = 30~\kms. The \oh\, line flux refers to the average flux of
      the two transitions of the $^2\Pi_{3/2}$ P4.5 (1+,1-) doublet. The
      FWHM and EW are the mean of the FWHM and EW of the two
      transitions. Total and column densities are computed assuming
      optically thin emission, LTE and T(OH) = T(H${\rm _2}$O) = 700\,K
      and T(OH) = T(H${\rm _2}$O) = 1000\,K respectively.}
    \tablenotetext{**}{In this case the \oh\, doublet is blended
      (Fig.~\ref{fig:oh3} \& \ref{fig:oh4}). The FWHM and EW are computed as
      the half of the total doublet FWHM and EW.}
  \end{deluxetable}

\end{document}